\documentstyle{article}
\begin{document}

\title{A.C. Conductivity of a Disordered Metal}
\author{ Girish S. Setlur \\
 The Institute of Mathematical Sciences \\ Chennai 600113, India. }
 
\maketitle                 

\begin{abstract}
 The degenerate free Fermi gas coupled to a random potential is used
 to compute a.c. conductivity in various dimensions.
 We first formally diagonalise the hamiltonian using an appropriate
 basis that is a functional of the disorder potential. Then we compute
 the a.c. conductivity at zero temperature using the Kubo formula.
 This a.c. conductivity is a functional of the disordered potential.
 The wavefunction of extended states is written as exponential of the
 logarithm. We use the cumulant expansion to compute the disordered
 averaged a.c. conductivity for Gaussian disorder.
 The formula is valid if a certain linearization approximation
 is valid in the long-wavelength limit. 
\end{abstract} 

\section{Introduction}

  A simple theory of the Anderson transition is presented that directly
  computes measurable quantities such as a.c. conductivity. 
  The relevant literature on this subject is vast 
  and we shall not attempt to be exhaustive in surveying it.
  Anderson's pioneering
   work on localization\cite{Anderson} was followed by the 
   work of Abrahams et.al. \cite{gangoffour}
   and later on a more rigorous formulation
   of the notion of disorder averaging was given by McKane and
   Stone\cite{McKane}.
   This relates to a single electron in a disordered potential. 
   The classic review of Lee and Ramakrishnan \cite{Lee}
   includes many references on the
   literature concerning the degenerate electron gas in a disordered 
   potential. A more recent review is by Abrahams and \cite{Abra}.
   Belitz and Kirkpatrick\cite{Kirk} have a review that is aimed at
   theorists. The present write-up is intended to give a simple derivation
   of the Anderson transition in three dimensions. 

 The basic idea of this work is as follows. We formally diagonalise the 
 hamiltonian of a single electron in a fixed disorder potential.
 The wavefuntion of the electron then may be formally written as,
\begin{equation}
\phi_{i}({\bf{x}}) = \frac{ C_{i} }{ \sqrt{V} }
e^{i \theta_{i}({\bf{x}}) } R_{i}^{\frac{1}{2}}({\bf{x}})
\end{equation}
 Here $ C_{i} $ is an appropriate normalization constant.
 The novel feature involves rewriting the square 
 of the amplitude $ R_{i}({\bf{x}}) $ also as an exponential.
\begin{equation}
R_{i}({\bf{x}}) = e^{ {\tilde{\Lambda}}_{i}({\bf{x}}) }
\end{equation}
 Then one uses the observation that only extented states participate in the
 conduction. Thus we may write,
\begin{equation}
\phi_{i}({\bf{x}}) = \frac{ C_{i} }{ \sqrt{V} }
e^{i {\bf{k}}_{i}.{\bf{x}} +
 i {\tilde{\theta}}_{i}({\bf{x}}) } e^{ \frac{1}{2} {\tilde{\Lambda}}_{i}({\bf{x}}) }
\label{EQNPHI}
\end{equation}
 Then we make the ansatz that
 $ <{\tilde{\theta}}_{i}>_{dis}  = <{\tilde{\Lambda}}_{i}>_{dis} = 0 $.
 This ansatz is shown to be consistent in the main text. 
 Then one uses the Kubo formula to evaluate the a.c.
 conductivity. One simply plugs in the form in Eq.(~\ref{EQNPHI}) 
 into this formula and uses the cumulant expansion to evluate the
 disorder averaged conductivity. In the cumulant expansion, we encounter
 correlation functions such as
 $  <{\tilde{\theta}}_{i}({\bf{x}})
 {\tilde{\theta}}_{j}({\bf{x}}^{'}) >_{dis} $, 
 $  <{\tilde{\theta}}_{i}({\bf{x}})
 {\tilde{\Lambda}}_{j}({\bf{x}}^{'}) >_{dis} $
 and so on. These have been explicitly computed in the appendix. 
 It can be shown that in some approximate sense, the higher order
 correlation functions all vanish. Hence we may deduce a
 formula for the disorder averaged a.c. conductivity provided it
 is legitimate to use only the linearization approximation.
 Our formalism does not 
 have an arbitrarily chosen cutoff. A natural smooth cutoff emerges
 from a careful treatment of curvature effects by retaining double
 derivatives, in other words, by not linearizing the bare fermion dispersion.
 In passing we note that this formalism involves no bosonization or
 other advanced field theoretic ideas. It only involves the
 simple Kubo formula. The answer for the a.c. conductivity may
 be written down as follows for Gaussian disorder.
 It has been reduced to quadratures.

\newpage

\begin{equation}
 Re[ \sigma(\omega) ] = 
\left( \frac{ \pi e^{2} }{mV} \right)
\frac{ 1 }{\omega}
\sum_{ {\bf{k}}_{i}, {\bf{k}}_{j} }
 \left< {\bf{J}}_{ji}(U_{dis}) \cdot {\bf{J}}_{ij}(U_{dis}) \right> 
\mbox{         }\theta(\epsilon_{F}-\frac{ {\bf{k}}^{2}_{j} }{2m} )  
\theta(\omega + \frac{ {\bf{k}}^{2}_{j} }{2m} -\epsilon_{F})
 \mbox{         }
 \delta(\omega -  \frac{ {\bf{k}}^{2}_{i} }{2m}
 + \frac{ {\bf{k}}^{2}_{j} }{2m} )
\end{equation}
where,
\begin{equation}
 \left< {\bf{J}}_{ji}(U_{dis}) \cdot {\bf{J}}_{ij}(U_{dis}) \right> 
 = -\frac{1}{V} \int d^{d} x \mbox{       }
e^{ i({\bf{k}}_{j}-{\bf{k}}_{i}).{\bf{x}} }
e^{ \frac{1}{2} <E^{'2}> } \mbox{      }
F({\bf{k}}_{i},{\bf{k}}_{j})
\end{equation}
\begin{equation}
\frac{1}{2} <E^{'2}> =
 - \frac{ m^{2} \Delta^{2} }{V} \sum_{ {\bf{q}} }
 \left( \frac{1}{ D_{i}({\bf{q}}) }  - \frac{1}{ D_{j}({\bf{q}}) } \right)^{2} 
+ \frac{ m^{2} \Delta^{2} }{V} \sum_{ {\bf{q}} }
\left( \frac{ 1 }{ D_{i}(-{\bf{q}}) }
 + \frac{ 1 }{ D_{j}({\bf{q}}) } \right)^{2}
e^{-i{\bf{q}}.{\bf{x}}}
\end{equation}
\[
F({\bf{k}}_{i},{\bf{k}}_{j}) = \frac{1}{2V} \sum_{ {\bf{q}} }\frac{2 m^{2} \Delta^{2} {\bf{q}}^{2} }
{ D_{i}(-{\bf{q}})D_{j}({\bf{q}})} 
e^{ -i{\bf{q}}.{\bf{x}} }
\]
\[
+ [ i{\bf{k}}_{i} 
 + \frac{1}{2V}
\sum_{ {\bf{q}} }
\frac{ 2 m^{2} \Delta^{2} }{ D_{i}({\bf{q}})D_{j}({\bf{q}}) }
 (-i{\bf{q}})
+ \frac{1}{2V}
\sum_{ {\bf{q}} }\left( 
\frac{ 2 m^{2} \Delta^{2} }{ D_{i}(-{\bf{q}})D_{j}({\bf{q}}) }
 (i{\bf{q}})
 + \frac{ 2 m^{2} \Delta^{2} }{ D^{2}_{i}(-{\bf{q}}) }
 (i{\bf{q}}) \right)
e^{ -i{\bf{q}}.{\bf{x}} } ]
\]
\begin{equation}
\cdot \mbox{     }
 [ i{\bf{k}}_{j} + \frac{1}{2V}
\sum_{ {\bf{q}} }
\frac{ 2 m^{2} \Delta^{2} }{ D_{i}({\bf{q}})D_{j}({\bf{q}}) }
 (-i{\bf{q}})
+ \frac{1}{2V}
\sum_{ {\bf{q}} }\left( 
\frac{ 2 m^{2} \Delta^{2} }{ D_{i}(-{\bf{q}})D_{j}({\bf{q}}) }
 (-i{\bf{q}})
 +  \frac{ 2 m^{2} \Delta^{2} }{ D^{2}_{j}({\bf{q}}) }
 (-i{\bf{q}}) \right)
e^{ -i{\bf{q}}.{\bf{x}} } ]
\end{equation}
\begin{equation}
D_{i}({\bf{q}}) = -{\bf{k}}_{i}.{\bf{q}} + \frac{ {\bf{q}}^{2} }{2}
\end{equation}
 The presence of the $ {\bf{q}}^{2}/2 $ makes
 the integrals finite at high momenta. 
 This comes about by retaining double derivaties,
 in other words by taking into account the parabolic nature of the band. 
The remaining sections explain how these formulas are derived.
Before we do this we would like to point out some general facts.
Since,
\begin{equation}
f({\bf{k}}_{i},{\bf{k}}_{j}) \equiv
 \left< {\bf{J}}_{ji}(U_{dis}) \cdot {\bf{J}}_{ij}(U_{dis}) \right> 
 \equiv  \left< |{\bf{J}}_{ij}(U_{dis})|^{2} \right> \geq 0
\end{equation}
We may sum over the angles first to write,
\begin{equation}
\frac{1}{V}{\tilde{f}}(|{\bf{k}}_{i}|,|{\bf{k}}_{j}|) \equiv
\sum_{ {\hat{k}}_{i}, {\hat{k}}_{j} }
f({\bf{k}}_{i},{\bf{k}}_{j}) \geq 0
\end{equation}

\newpage

\begin{equation}
 Re[ \sigma(\omega) ] = 
\left( \frac{ \pi e^{2} }{m} \right)
\frac{ 1 }{\omega}
\int_{0}^{\infty} dk_{i} \mbox{     } k^{d-1}_{i} \mbox{      }
 \int_{0}^{\infty} dk_{j} \mbox{     } k^{d-1}_{j} \mbox{      }
  {\tilde{f}}(k_{i},k_{j})
\mbox{         }\theta(\epsilon_{F}-\frac{ k^{2}_{j} }{2m} )  
\theta(\omega + \frac{ k^{2}_{j} }{2m} -\epsilon_{F})
 \mbox{         }
 \delta(\omega -  \frac{ k^{2}_{i} }{2m}
 + \frac{ k^{2}_{j} }{2m} )
\end{equation}
For $ m\omega \ll k^{2}_{F} $, we have,
\begin{equation}
 Re[ \sigma(\omega) ] 
\sim \mbox{           }
  {\tilde{f}}(k_{F}(1 + m\omega/k^{2}_{F}),k_{F})
\end{equation}
 One really important question that we have to face is whether the 
 zero frequency limit of the a.c. conductivity (obtained via
 Kubo formula) is the d.c. conductivity.
 According to the review by Lee and Ramakrishnan\cite{Lee} 
 this is true only in three dimensions.
 In one and two dimensions, it appears that according to
 the results presented,
 the d.c. conductivity may not be
 obtained by taking the zero frequency limit of the a.c. conductivity.
 They show that the a.c. conductivity diverges logarithimically for small 
 frequencies in two dimensions whereas the d.c. conductivity is strictly zero. 
 A frequency cutoff emerges in these dimensions that tells us
 that for frequencies smaller than this cutoff the formulas
 given there break down. According to the analysis given
 in Appendix C of
 this article, there should be no problem in any number of dimensions.
 The zero frequency limit of the a.c. conductivity is in fact the d.c.
 conductivity.
 In three dimensions we may expect to find the
 mobility edge while computing the d.c. conductivity.
\begin{equation}
 \sigma_{d.c.} 
\sim \mbox{           }
  {\tilde{f}}(k_{F},k_{F})
\end{equation}

{\bf{Results So Far}} :

 So far the results have been very disappointing. In one dimension, we expect
 to see that the d.c. conductivity is zero. Instead we find that it is not
 although there appear to be many subtle cancellations. It is likely that
 the linearization is not adequate and retaining some nonlinear terms may be
 required.
 The main culprit seems to be an approximation that replaces a certain variable
 quadratic in the $ {\tilde{\theta}} $ and $ {\tilde{\Lambda}} $ by it disorder
 average in an effort to render the equations linear differential equations.
 However when evaluated this average is zero and hence it is a really 
 bad idea. Hwoever the author had hoped that it would not be serious.
 But it is. Future attempts may try and address these problems,
 although there are many good ideas in this preprint. 

\section{Some Technical Musings}

 It appears that the mathematical literature on the subject of 
 quantum particles in random potentials is vast\cite{Krishna}.  
 It is possible, indeed likely that many mathematically rigorous 
 results are known regarding this problem. But this does not prevent
 the authors from making some remarks that more knowledgeable
 readers may choose to critique. In particular, the author is
 uncomfortable with the notion of disorder averaging. Nature
 chooses its potentials based on the distribution of impurities, defects and
 so on. This potential is fixed and well-defined for a particular distribution
 of these imperfections. The physicists' ignorance of the precise nature
 of this potential is not a license to average over these potentials. 
 Nature does not average, people do. But are people justified in averaging ?
 In other words can averaging simplify the problem without washing out
 essential physics ? In order to answer this question we have to make
 the following conjectures. 

\vspace{0.1in}

\noindent {\bf{Defn0}} : Let $ {\mathcal{U}}_{d} $ be the set of
 all potentials $ U({\bf{x}}) $ in a fixed spatial dimension $ d $.

\noindent {\bf{Defn1}} : Let $ {\mathcal{F}}_{d} $ be the set of
 all potentials $ U({\bf{x}}) $ in a fixed spatial dimension $ d $
 that has the following property. They all lead to the same
 exponent $ \delta $ for the frequency dependence of the a.c. conductivity.
 In other words, each of these potentials predicits that
 $ Re[ \sigma(\omega) ] \sim \omega^{\delta} $
 (in some region of $ \omega $ with possibly
 some additive part independent of $ \omega $)
  with the {\it{same}} $ \delta $.

\noindent {\bf{Conjecture 1}} :  $ {\mathcal{F}}_{d} $ is dense in
 $ {\mathcal{U}}_{d} $. 

 If $ {\bf{Conjecture 1}} $ is valid, then one may average over all these
 `sufficiently erratic' potentials and expect to extract $ \delta $ which
 is all that physicists care about. It is possible that $ \delta $ may
 be extracted from a numerical solution of the Schrodinger equation 
  using a specific $ U $ that 
 belongs to the set $  {\mathcal{F}}_{d} $. But this would involve
 using the computer for more than checking one's email, 
 and not everyone likes that.

\noindent {\bf{Defn2}} : Let $ {\mathcal{M}}_{3} $ be the set of
 all potentials $ U({\bf{x}}) $ in spatial dimension $ d = 3 $
 that has the following property. They all lead to the same
 exponent $ \beta $ for the mobility edge exponent.
 In other words, each of these potentials predicit that
 $ \sigma_{d.c.} \sim (E_{F} - E_{c})^{\beta}
 \mbox{     }\theta(E_{F}-E_{c}) $ with the {\it{same}} $ \beta $.
 However for different potentials, $ E_{c} $ - the mobility
 edge, may be different.

\noindent {\bf{Conjecture 2}} :  $ {\mathcal{F}}_{3} $ is dense in
 $ {\mathcal{U}}_{3} $. 

 If $ {\bf{Conjecture 2}} $ is valid, then one may average over all these
 `sufficiently erratic' potentials and expect to extract $ \beta $.

 Thus the validity of the process of averaging over potentials rests crucially
 it seems, on all these sufficiently erratic potentials predicting the same
 exponents and on these sufficiently erratic potentials spanning nearly all
 possible potentials. 

 If both these are satistifed then one may average over all potentials and
 extract the exponents, or, if one is better at programming, 
 choose a particular potential from this set, numerically solve the
 Schrodinger equation and extract the exponent from there. In either
 case we should get the same answer. A final conjecture seems appropriate.

 {\bf{ Conjecture 3 }} : Let $ {\mathcal{M}}^{'}_{3} $ have an exponent
 $ \beta^{'} $ and  $ {\mathcal{F}}^{'}_{d} $ have an exponent $ \delta^{'} $,
 then $ \beta = \beta^{'} $ and $ \delta = \delta^{'} $. In other words,
 these exponents are unique.

 With powerful computers now available, purely analytical methods such as this
 work may seem  pass\`{e},
 but a closed formula for the a.c. conductivity
 that one can stare at (and one that is hopefully right) and admire
 has a charm that a cold data file on the hard disk is unable to 
 duplicate. Besides, with Coulomb interaction, the problem becomes intractable
 numerically, however, one may expect to combine the sea-boson method with
 the present one to extract the exponents analytically. 

\section{A.C. Conductivity Using Kubo Formula}  
  
 In this section, we derive a formula for the a.c. conductivity 
 in terms of the total momentum-momentum correlation function.
 To derive this, observe that the relevant
 hamiltonian that couples to external fields is of the form,
\begin{equation}
H_{ext}(t) = -\frac{ |e| }{m} {\bf{A}}_{ext}(t) \cdot {\bf{P}}
\end{equation}
where $ {\bf{P}} = \sum_{ {\bf{k}} }{\bf{k}} \mbox{         }
c^{\dagger}_{ {\bf{k}} }c_{ {\bf{k}} } $ is the total momentum operator.
 Thus we may define formally the net momentum in the presence of the 
 external field in the imaginary time formalism in the
 interaction representation as
\begin{equation}
\left< {\bf{P}}(t) \right>
 = \frac{ \left< T \mbox{   }S\mbox{        }{\hat{ {\bf{P}} }}(t) 
 \right> }{ \left< T \mbox{   }S\mbox{        } \right> }
\end{equation}
where the S-matrix is defined as,
\begin{equation}
S = e^{i\int_{0}^{-i\beta} dt \frac{ |e| }{m} {\bf{A}}_{ext}(t)
 \cdot {\hat{ {\bf{P}} }}(t) }
\end{equation}
 Here $ {\hat{ {\bf{P}} }}(t) $ evolves according to the
 time-independent hamiltonian. The net current density in the system is
 given by $ {\bf{J}}(t) = (|e|/mV) \mbox{         }{\bf{P}}(t) $,
 this is in units of charge flowing per unit area per unit time
 in three space dimensions.
 From Ohm's law we expect,
\begin{equation}
\left< {\bf{J}}(t) \right>
 = \int_{0}^{-i\beta} dt^{'} \mbox{          }
{\tilde{\sigma}}(t-t^{'}) \mbox{       }
{\bf{E}}_{ext}(t^{'}) 
\end{equation}
 The a.c. conductivity for complex frequencies is then given by,
\begin{equation}
\sigma(iz) = \int_{0}^{-i\beta} dt \mbox{          }{\tilde{\sigma}}(t) 
\mbox{         }e^{-z\mbox{   } t}
\end{equation}
The d.c. conductivity is then given by,
\begin{equation}
\sigma_{d.c.} = \sigma(i0 + \epsilon)
\end{equation}
where, $ \epsilon = 0^{+} $.
\begin{equation}
{\tilde{\sigma}}(t-t^{'}) = \frac{ |e| }{mV}
 \mbox{         }
 \left( \frac{ \delta \left< {\bf{P}}(t) \right> }
{ \delta {\bf{E}}_{ext}(t^{'}) } 
 \right)_{ {\bf{E}}_{ext} \equiv 0 }
\end{equation}
\begin{equation}
{\tilde{\sigma}}(t-t^{'}) = \frac{ |e| }{mV}
 \mbox{         }
\frac{ \delta }{ \delta {\bf{E}}_{ext}(t^{'}) }
\frac{ \left< T \mbox{   }S\mbox{        }{\hat{ {\bf{P}} }}(t) 
 \right> }{ \left< T \mbox{   }S\mbox{        } \right> }
\end{equation}
Since $ {\bf{E}}_{ext}(t) = -\partial {\bf{A}}_{ext}(t) /\partial t $, we have,
\begin{equation}
\frac{ \delta {\bf{A}}_{ext}(t) }{ \delta {\bf{E}}_{ext}(t^{'}) } 
 = -\theta(t-t^{'})
\end{equation}
\begin{equation}
\frac{ \delta }{ \delta {\bf{E}}_{ext}(t^{'}) }
S = S \left( -i \int_{0}^{-i\beta}
dt_{1} \mbox{       }\theta(t^{'}-t_{1})
\frac{ |e| }{m}
{\hat{ {\bf{P}} }}(t_{1}) 
\right)
\end{equation}
\begin{equation}
{\tilde{\sigma}}(t-t^{'}) = -\frac{ i \mbox{  }e^{2} }{m^{2}V}
 \mbox{         }
\int_{0}^{-i\beta} dt_{1}
\theta(t^{'}-t_{1})
\left< T \delta {\hat{ {\bf{P}} }}(t_{1}) \cdot 
\delta {\hat{ {\bf{P}} }}(t) \right>
\end{equation}
\begin{equation}
\left< T \delta {\hat{ {\bf{P}} }}(t_{1}) \cdot 
\delta {\hat{ {\bf{P}} }}(t) \right>
 = \theta(t_{1}-t) \mbox{       }
\sum_{ {\bf{k}} {\bf{k}}^{'} }({\bf{k}} \cdot {\bf{k}}^{'})
N({\bf{k}},t_{1}-t;{\bf{k}}^{'},0)
 +  \theta(t-t_{1}) \mbox{       }
\sum_{ {\bf{k}} {\bf{k}}^{'} }({\bf{k}} \cdot {\bf{k}}^{'})
N({\bf{k}},t-t_{1};{\bf{k}}^{'},0)
\end{equation}
If we write for the dynamical number-number correlation function,
\begin{equation}
N({\bf{k}},t;{\bf{k}}^{'},0)
 \equiv \left< n_{ {\bf{k}} }(t) n_{ {\bf{k}}^{'} }(0) \right>
 - \left< n_{ {\bf{k}} }(t) \right> \left< n_{ {\bf{k}}^{'} }(0) \right>
 = \sum_{ij} 
e^{-i(\epsilon_{i}-\epsilon_{j})t}
{\tilde{N}}({\bf{k}},\epsilon_{i},\epsilon_{j};{\bf{k}}^{'},0)
\end{equation}
where $ n_{ {\bf{k}} } = c^{\dagger}_{ {\bf{k}} }c_{ {\bf{k}} } $.
We may write,
\[
\left< T \delta {\hat{ {\bf{P}} }}(t_{1}) \cdot 
\delta {\hat{ {\bf{P}} }}(t) \right>
 = \theta(t_{1}-t) \mbox{       }
\sum_{ {\bf{k}} {\bf{k}}^{'} }({\bf{k}} \cdot {\bf{k}}^{'})
\sum_{ij} {\tilde{N}}({\bf{k}},\epsilon_{i},\epsilon_{j};{\bf{k}}^{'},0)
e^{-i(\epsilon_{i}-\epsilon_{j})(t_{1}-t)}
\]
\begin{equation}
 +  \theta(t-t_{1}) \mbox{       }
\sum_{ {\bf{k}} {\bf{k}}^{'} }({\bf{k}} \cdot {\bf{k}}^{'})
\sum_{i,j} {\tilde{N}}({\bf{k}},\epsilon_{i},\epsilon_{j};{\bf{k}}^{'},0)
e^{-i(\epsilon_{i}-\epsilon_{j})(t-t_{1})}
\end{equation}
 From the above formulas retaining only the terms that do not violate  
 causality and using $ iz_{n} \rightarrow \omega- i0^{+} $, we have
 at absolute zero,
\begin{equation}
Re[ \sigma(\omega,U_{dis}) ] = 
\left( \frac{ \pi e^{2} }{m^{2}V} \right)
\frac{ 1 }{\omega}
\sum_{ {\bf{k}} {\bf{k}}^{'} }
\sum_{i,j}
({\bf{k.k^{'}}})
 \mbox{         }{\tilde{N}}({\bf{k}},\epsilon_{i},\epsilon_{j};{\bf{k}}^{'},0)
 \delta(\omega - \epsilon_{i} + \epsilon_{j} )
\end{equation}
To compute this we have to first evaluate the 
dynamical number-number correlation function.
Consider the hamiltonian,
\begin{equation}
H = \sum_{ {\bf{k}} }\epsilon_{ {\bf{k}} }c^{\dagger}_{ {\bf{k}} }
c_{ {\bf{k}} }
 + \sum_{ {\bf{q}} } \frac{ U_{dis}({\bf{q}}) }{\sqrt{V}}
\sum_{ {\bf{k}} }c^{\dagger}_{ {\bf{k}} + {\bf{q}}/2  }
c_{ {\bf{k}} - {\bf{q}}/2 }
\end{equation}
This may be diagonalised by the following formal transformation.
\begin{equation}
c_{ {\bf{k}} } = \sum_{i} \varphi_{i}({\bf{k}})d_{i}
\end{equation}
\begin{equation}
<c^{\dagger}_{ {\bf{k}} }c_{ {\bf{k}} }>
 = \sum_{i} |\varphi_{i}({\bf{k}})|^{2} \theta(\epsilon_{F}-\epsilon_{i})
\end{equation}
\begin{equation}
n_{ {\bf{k}} }(t) =
 \sum_{ij} 
\varphi^{*}_{j}({\bf{k}}) 
\varphi_{i}({\bf{k}})d^{\dagger}_{j}d_{i}
e^{-i(\epsilon_{i}-\epsilon_{j})t} 
\end{equation}
\begin{equation}
n_{ {\bf{k}}^{'} }(0) =
 \sum_{i^{'}j^{'}} 
\varphi^{*}_{j^{'}}({\bf{k}}^{'}) 
\varphi_{i^{'}}({\bf{k}}^{'})d^{\dagger}_{j^{'}}d_{i^{'}}
\end{equation}
\begin{equation}
N({\bf{k}},t;{\bf{k}}^{'},0)
 =\sum_{i,j} \varphi^{*}_{j}({\bf{k}}) 
\varphi_{i}({\bf{k}})
\varphi^{*}_{i}({\bf{k}}^{'}) 
\varphi_{j}({\bf{k}}^{'})
\theta(\epsilon_{F}-\epsilon_{j})  
\theta(\epsilon_{i}-\epsilon_{F})
e^{-i(\epsilon_{i}-\epsilon_{j})t} 
\end{equation}
This means,
\begin{equation}
\sum_{ {\bf{k}} {\bf{k}}^{'} }
({\bf{k.k^{'}}}) \mbox{       }
{\tilde{N}}({\bf{k}},\epsilon_{i},\epsilon_{j};{\bf{k}}^{'},0)
 = 
{\bf{J}}_{ji}(U_{dis}) \cdot {\bf{J}}_{ij}(U_{dis}) 
\theta(\epsilon_{F}-\epsilon_{j})  
\theta(\epsilon_{i}-\epsilon_{F})
\end{equation}
 Here $ \phi_{i}({\bf{x}}) $ is the solution to the equation below, and
 $ \varphi_{i}({\bf{k}}) $ is its Fourier  transform.
\begin{equation}
\left( - \frac{ \nabla^{2} }{2m} + U_{dis}({\bf{x}}) \right) 
  \phi_{i}({\bf{x}}) = \epsilon_{i} \mbox{        }
\phi_{i}({\bf{x}})
\end{equation}
\begin{equation}
 {\bf{J}}_{ij}(U_{dis}) = \sum_{ {\bf{k}} }{\bf{k}} \mbox{        }
\varphi^{*}_{i}({\bf{k}}) \varphi_{j}({\bf{k}}) 
= -\int d^{d}x \mbox{         }
\phi^{*}_{i}({\bf{x}}) i\nabla_{ {\bf{x}} }\phi_{j}({\bf{x}}) 
\end{equation}
\begin{equation}
Re[ \sigma(\omega,U_{dis}) ] = 
\left( \frac{ \pi e^{2} }{mV} \right)
\frac{ 1 }{\omega}
\sum_{i,j}{\bf{J}}_{ji}(U_{dis}) \cdot {\bf{J}}_{ij}(U_{dis}) 
\mbox{         }\theta(\epsilon_{F}-\epsilon_{j})  
\theta(\omega + \epsilon_{j}-\epsilon_{F})
 \mbox{         }
 \delta(\omega - \epsilon_{i} + \epsilon_{j} )
\label{SIGMADIS}
\end{equation}
 Till now the discussion has been at the formal level.
 Now we would like to compute the disorder averaged conductivity
 assuming a Gaussian disorder. In other words, the averages over the
 disordered potential have to be performed using the conditions,
\begin{equation}
<U_{dis}({\bf{x}})> = 0
\end{equation}
\begin{equation}
<U_{dis}({\bf{x}}) U_{dis}({\bf{x}}^{'})> = \Delta^{2}
 \mbox{        }\delta^{d}({\bf{x}}-{\bf{x}}^{'})
\end{equation}
 Thus the sum over all configuarations of disordered potential keeps
 the sum $ \int d^{d}x \mbox{        }U_{dis}^{2}({\bf{x}}) $
 fixed and sums the conductivity
 obtained from each configuration and then computes the average. 
 Several observations may be made regarding this. First, the
 energy $ \epsilon_{i} $ for each choice of $ U_{dis} $ may be
 discrete and negative (bound state) or positive and continuous, corresponding
 to Bloch waves. The delta-function forces $ \omega $ to be equal
 to the difference $ \epsilon_{i} - \epsilon_{j} $. 
 For small $ \omega $, the difference $ \epsilon_{i} - \epsilon_{j} $  
 is likely to be comparable to $ \omega $ only for the Bloch states.
 That is to say, that only the extended states participate in conduction.
 Thus we may polar decompose the wavefunction $ \phi_{i} $ and assume that
 the magnitude is (roughly) independent of position,
 a feature characteristic of extended states. 
 We have found that these simplifying assumptions though natural
 lead to some divergences in the ultraviolet regime.
 Thus we shall have to retain the double derivatives. This 
 is done in the appendix and the correlations between
 the $ {\tilde{\Lambda}}'s $ and the $ {\tilde{ \theta }}'s $(see below)
  and amongst themselves
 are derived. 
Thus we write,
\begin{equation}
\phi_{i}({\bf{x}}) = e^{ i\theta_{i}({\bf{x}}) }
R_{i}^{\frac{1}{2}}({\bf{x}})
\end{equation}
 For some phase $ \theta_{i} $
 and $ R_{i} $ that is a function of the disordered potential.
          These obey the following set of coupled equations.
\begin{equation}
-\frac{ \nabla^{2}  \theta_{i} }{2m}
 -\frac{ \nabla \theta_{i} \cdot \nabla R_{i} }{2m R_{i}} = 0
\end{equation}
\begin{equation}
\frac{ (\nabla \theta_{i})^{2} }{2m} 
 - \frac{1}{8m} \left[ \frac{ 2 \nabla^{2} R_{i} }{R_{i}}
 - \frac{ (\nabla R_{i})^{2} }{R_{i}^{2}} \right]
 = \epsilon_{i} - U_{dis}
\end{equation}
Now we set,
\begin{equation}
R_{i}({\bf{x}}) = \frac{ C^{0}_{i} }{ V } \mbox{           }
e^{ {\tilde{\Lambda}}_{i}({\bf{x}}) }
\end{equation}
 Here $  C^{0}_{i} $ is a normalization constant.
 It may be determined as follows. 
\begin{equation}
\left< \int d^{d}x \mbox{      }
\left| \phi_{i}({\bf{x}}) \right|^{2} \right>_{dis} = 
\int d^{d}x \mbox{      } \left< R_{i}({\bf{x}}) \right>_{dis}
 = \frac{ C^{0}_{i} }{ V } \mbox{     }
\int d^{d}x \mbox{      }
\left< e^{ {\tilde{\Lambda}}_{i}({\bf{x}}) } \right>
 = C^{0}_{i} \mbox{           }e^{ \frac{1}{2} 
\left< {\tilde{\Lambda}}^{2}_{i}({\bf{x}}) \right> } = 1 
\end{equation}
Thus,
\begin{equation}
C^{0}_{i} = e^{ -\frac{1}{2} 
\left< {\tilde{\Lambda}}^{2}_{i}({\bf{x}}) \right>_{dis} }
\end{equation}
 It will be shown in the appendices
 that $ \left< {\tilde{\Lambda}}^{2}_{i}({\bf{x}}) \right> $
 is independent of position.
 Further since only extended
 states participate in the conduction, we may
 write $ \theta_{i}({\bf{x}}) = {\bf{k}}_{i} \cdot {\bf{x}} + {\tilde{\theta}}_{i}({\bf{x}}) $. 
 In the appendices, an approximate
 scheme is written down that allows for a relatively simple
 computation of the average
 {\mbox{ $ < {\bf{J}}_{ij}(U_{dis}) \cdot {\bf{J}}_{ji}(U_{dis})> $ }}. 
 From Eq.(~\ref{SIGMADIS}) it is clear that we would very much
 like to consider the eigenenergies $ \epsilon_{i} $ as being
 independent of $ U_{dis} $ in the sense that we may replace
 {\mbox{ $ \epsilon_{i} \equiv <\epsilon_{i}>_{dis} $ }}.
 This requires some justification. Also from the appendix we see that
\begin{equation}
 \nabla^{2}  {\tilde{\theta}}_{i} 
 + {\bf{k}}_{i} \cdot \nabla {\tilde{\Lambda}}_{i} 
 + \nabla {\tilde{\theta}}_{i} \cdot \nabla {\tilde{\Lambda}}_{i} 
 = 0
\label{EQN1}
\end{equation}
\begin{equation}
\frac{ {\bf{k}}^{2}_{i} }{2m}
 + \frac{ (\nabla {\tilde{\theta}}_{i})^{2} }{2m} 
+ \frac{ {\bf{k}}_{i} \cdot \nabla {\tilde{\theta}}_{i} }{m}
 - \frac{1}{8m} \left[ 2 \nabla^{2} {\tilde{\Lambda}}_{i}
 +  (\nabla {\tilde{\Lambda}}_{i})^{2} \right]
 = \epsilon_{i} - U_{dis}
\label{EQN2}
\end{equation} 
 The above Eq.(~\ref{EQN1}) and  Eq.(~\ref{EQN2}) are absolutely exact.
 The approximations arise when we decide to 
 linearize the above nonlinear partial differential equations by
 replacing the quadratic parts with their disorder averages.
 Obviously, this is justifiable only if we show that
 the fluctuations of the operators we have averaged out are small,
 preferably zero. First we try and justify 
 $ \epsilon_{i} \equiv <\epsilon_{i}>_{dis} $. For this we have to evaluate
 $ <( \epsilon_{i} -  <\epsilon_{i}>_{dis})^{2}>^{\frac{1}{2}} $
 and show that it is small compared to $  <\epsilon_{i}>_{dis} $.
 To do this we take recourse to perturbation theory. Using
 elementary perturbation theory upto second order we may write,
\begin{equation}
\epsilon_{i} = \frac{ {\bf{k}}^{2}_{i} }{2m} 
 + \int \frac{ d^{d}x }{V} \mbox{         }U_{dis}({\bf{x}})
 + \frac{1}{ V^{2} }\sum_{ {\bf{k}}_{j} }
\frac{1}{   \frac{ {\bf{k}}^{2}_{i} }{2m} 
 -  \frac{ {\bf{k}}_{j}^{2} }{2m} }
\left| \int d^{d}x \mbox{         }
e^{i({\bf{k}}_{i}-{\bf{k}}_{j}).{\bf{x}} }
U_{dis}({\bf{x}}) \right|^{2}
\end{equation}
Now,
\begin{equation}
< \epsilon_{i} >_{dis} = \frac{ {\bf{k}}^{2}_{i} }{2m} 
 + \frac{ \Delta^{2} }{ V }\sum_{ {\bf{k}}_{j} \neq {\bf{k}}_{i} }
\frac{1}{   \frac{ {\bf{k}}^{2}_{i} }{2m} 
 -  \frac{ {\bf{k}}_{j}^{2} }{2m} }
\label{AVEN}
\end{equation}
Therefore,
\begin{equation}
< \epsilon^{2}_{i} >_{dis} - < \epsilon_{i} >_{dis}^{2}
= \frac{ \Delta^{2} }{V} 
 + \frac{ \Delta^{4} }{ V^{2} }\sum^{'}_{ {\bf{k}}_{j} }
\frac{1}{   \left( \frac{ {\bf{k}}^{2}_{i} }{2m} 
 -  \frac{ {\bf{k}}_{j}^{2} }{2m} \right) }
 \mbox{         }
\frac{1}{ \left( \frac{ {\bf{k}}^{2}_{i} }{2m} 
 -  \frac{ (2{\bf{k}}_{i} - {\bf{k}}_{j})^{2} }{2m} \right) } 
 + \frac{ \Delta^{4} }{ V^{2} }\sum^{'}_{ {\bf{k}}_{j} }
\frac{1}{ \left( \frac{ {\bf{k}}^{2}_{i} }{2m} 
 -  \frac{ {\bf{k}}_{j}^{2} }{2m}  \right)^{2} }
\label{FLUC}
\end{equation}
 The sums over $ {\bf{k}}_{j} $ in Eq.(~\ref{FLUC}) are finite
 in all three dimensions and hence the fluctuation $ \epsilon_{i} $ is 
 vanishingly small in the thermodynamic limit. Even otherwise, within
 this scheme the sum in Eq.(~\ref{AVEN}) diverges, hence
 the fluctuation
 {\mbox{ $ (< \epsilon^{2}_{i} >_{dis} - < \epsilon_{i} >_{dis}^{2})^{\frac{1}{2}} $ }}  is bound to be small compared to the mean.
 The proof that we may legitimately linearize the equations 
 above (Eq.(~\ref{EQN1}) and Eq.(~\ref{EQN2})) is given below.
 First we note that two random variables $ O_{1}({\bf{x}}) $ 
 and $ O_{2}({\bf{x}}) $ may be replaced by their means
 if,
\begin{equation}
< O_{i}({\bf{x}}) O_{j}({\bf{x}}^{'}) > 
 = < O_{i}({\bf{x}}) >< O_{j}({\bf{x}}^{'}) >
\end{equation}
for $ i,j = 1,2 $.
 Thus we would like to make the following identifications.
\begin{equation}
\nabla {\tilde{\theta}}_{i} \cdot \nabla {\tilde{\Lambda}}_{i} 
 \approx \left< \nabla {\tilde{\theta}}_{i} \cdot
 \nabla {\tilde{\Lambda}}_{i} \right>_{dis} = 0 \mbox{  }
(from \mbox{   } appendix)
\label{CON1}
\end{equation}
\begin{equation}
\frac{ (\nabla {\tilde{\theta}}_{i})^{2} }{2m} 
 - \frac{1}{8m} (\nabla {\tilde{\Lambda}}_{i})^{2}
 \approx \frac{ < (\nabla {\tilde{\theta}}_{i})^{2} >_{dis}  }{2m} 
 - \frac{1}{8m} < (\nabla {\tilde{\Lambda}}_{i})^{2} >_{dis}
\label{CON2}
\end{equation}
 The first condition is satified if we ensure,
\begin{equation}
\left< \left( \nabla_{ {\bf{x}} }
 {\tilde{\theta}}_{i} \cdot \nabla_{ {\bf{x}} }
 {\tilde{\Lambda}}_{i} \right)
 \left( \nabla_{ {\bf{x}}^{'} }
 {\tilde{\theta}}_{j} \cdot \nabla_{ {\bf{x}}^{'} }
 {\tilde{\Lambda}}_{j} \right) \right>
 \approx 0
\end{equation}
In other words,
\[
\nabla^{m}_{ {\bf{x}} }  \nabla^{n}_{ {\bf{x}}^{'} }
 < {\tilde{\theta}}_{i}({\bf{x}}) {\tilde{\theta}}_{j}({\bf{x}}^{'}) >_{dis}
 \nabla^{m}_{ {\bf{x}} } \nabla^{n}_{ {\bf{x}}^{'} }
 < {\tilde{\Lambda}}_{i}({\bf{x}}) {\tilde{\Lambda}}_{j}({\bf{x}}^{'}) >_{dis}
\]
\begin{equation}
+ \nabla^{m}_{ {\bf{x}} }  \nabla^{n}_{ {\bf{x}}^{'} }
 < {\tilde{\theta}}_{i}({\bf{x}})  {\tilde{\Lambda}}_{j}({\bf{x}}^{'}) >_{dis}
\nabla^{m}_{ {\bf{x}} } \nabla^{n}_{ {\bf{x}}^{'} }
  < {\tilde{\Lambda}}_{i}({\bf{x}}) {\tilde{\theta}}_{j}({\bf{x}}^{'}) >_{dis}
 \approx 0
\label{EQNC1}
\end{equation}
The second condition is obeyed if we ensure,
\[
[ 2 \delta (\nabla_{ {\bf{x}} }  {\tilde{\theta}}_{i})^{2} 
 - \frac{1}{2} \delta 
 (\nabla_{ {\bf{x}} } {\tilde{\Lambda}}_{i})^{2} ] \mbox{   }
[ 2 \delta (\nabla_{ {\bf{x}}^{'} }  {\tilde{\theta}}_{j})^{2} 
 - \frac{1}{2}
\delta (\nabla_{ {\bf{x}}^{'} }  {\tilde{\Lambda}}_{j})^{2} ]
\]
\[
 = 8
 \left( \nabla^{m}_{ {\bf{x}} }  \mbox{     } \nabla^{n}_{ {\bf{x}}^{'} }
\left< {\tilde{\theta}}_{i}({\bf{x}})
 {\tilde{\theta}}_{j}({\bf{x}}^{'}) \right> \right)
 \left( \nabla^{m}_{ {\bf{x}} }  \mbox{    }\nabla^{n}_{ {\bf{x}}^{'} }
\left< {\tilde{\theta}}_{i}({\bf{x}})
 {\tilde{\theta}}_{j}({\bf{x}}^{'}) \right> \right)
\]
\[
+ \frac{1}{2} \left( \nabla^{m}_{ {\bf{x}} } \nabla^{n}_{ {\bf{x}}^{'} }
\left< {\tilde{\Lambda}}_{i}({\bf{x}})
 {\tilde{\Lambda}}_{j}({\bf{x}}^{'}) \right> \right)
\left( \nabla^{m}_{ {\bf{x}} } \nabla^{n}_{ {\bf{x}}^{'} }
\left< {\tilde{\Lambda}}_{i}({\bf{x}})
 {\tilde{\Lambda}}_{j}({\bf{x}}^{'}) \right> \right)
\]
\[
- 2 \left( \nabla^{m}_{ {\bf{x}} }  \nabla^{n}_{ {\bf{x}}^{'} }
 \left< {\tilde{\Lambda}}_{i}({\bf{x}})
 {\tilde{\theta}}_{j}({\bf{x}}^{'}) \right> \right)
\left( \nabla^{m}_{ {\bf{x}} } \nabla^{n}_{ {\bf{x}}^{'} }
 \left< {\tilde{\Lambda}}_{i}({\bf{x}})
 {\tilde{\theta}}_{j}({\bf{x}}^{'}) \right> \right)
\]
\begin{equation}
- 2 \left( \nabla^{m}_{ {\bf{x}} } \nabla^{n}_{ {\bf{x}}^{'} }
\left< {\tilde{\theta}}_{i}({\bf{x}}) 
 {\tilde{\Lambda}}_{j}({\bf{x}}^{'}) \right> \right)
\left( \nabla^{m}_{ {\bf{x}} } \nabla^{n}_{ {\bf{x}}^{'} }
\left< {\tilde{\theta}}_{i}({\bf{x}}) 
 {\tilde{\Lambda}}_{j}({\bf{x}}^{'}) \right> \right)
 \approx 0
\label{EQNC2}
\end{equation}
 Finally, we have to also ensure that the cross correlation functions are zero.
\[
\frac{1}{m}
\left( \nabla^{ m }_{ {\bf{x}} }  \nabla^{ n }_{ {\bf{x}}^{'} } 
< {\tilde{\theta}}_{i}({\bf{x}}) {\tilde{\theta}}_{j}({\bf{x}}^{'}) > \right)
\left(  \nabla^{ m }_{ {\bf{x}} }  \nabla^{ n }_{ {\bf{x}}^{'} } 
  \left< {\tilde{\Lambda}}_{i}({\bf{x}}) 
{\tilde{\theta}}_{j}({\bf{x}}^{'}) \right>  \right)
\]
\begin{equation}
 - \frac{1}{4m} 
 \left( \nabla^{ m }_{ {\bf{x}} }  \nabla^{ n }_{ {\bf{x}}^{'} } 
< {\tilde{\theta}}_{i}({\bf{x}}) {\tilde{\Lambda}}_{j}({\bf{x}}^{'}) > \right)
\left(  \nabla^{ m }_{ {\bf{x}} }  \nabla^{ n }_{ {\bf{x}}^{'} } 
  \left< {\tilde{\Lambda}}_{i}({\bf{x}}) 
{\tilde{\Lambda}}_{j}({\bf{x}}^{'}) \right>  \right)
 \approx 0
\label{EQNC3}
\end{equation}
  The three Eq.(~\ref{EQNC1}), Eq.(~\ref{EQNC2}) and  Eq.(~\ref{EQNC3})
  represent correlation functions
  of random variables that we have assumed may be replaced by their averages.
  Unfortunately, we have found that these conditions are too
 severe to be obeyed. Thus we have to be content at saying that the formulas 
 are not exact, but are based on a plausible linearization assumption. 
 This assumption is `self-consistent' in the sense that no obvious  
 inconsistency shows up.
 A hand-waving justification for at least Eq.(~\ref{CON2}) is that the
 right hand side diverges when evaluated thus the fluctuation of
 this quantity being finite is certainly small compared to the
 mean which is formally infinite.
 This infinity may be absorbed by a suitable redefinition of the Fermi energy.
 On the other hand, Eq.(~\ref{CON1}) is impossible to justify except
 that making it renders an elegant
 analytical solution possible since the equations are now linear. 
 
\newpage

\section{Appendix A}

\begin{equation}
\frac{ \nabla R_{i} }{R_{i}}
 = \nabla Ln[R_{i}] 
 =  \nabla {\tilde{\Lambda}}_{i}
\end{equation}
We then have to solve,
\begin{equation}
 \nabla^{2}  \theta_{i} 
 + \nabla \theta_{i} \cdot \nabla {\tilde{\Lambda}}_{i}  = 0
\end{equation}
\begin{equation}
\frac{ (\nabla \theta_{i})^{2} }{2m} 
 - \frac{1}{8m} \left[ 2 \nabla^{2} {\tilde{\Lambda}}_{i}
 +  (\nabla {\tilde{\Lambda}}_{i})^{2} \right]
 = \epsilon_{i} - U_{dis}
\end{equation}
 We now decompose $ \theta_{i} $ as follows.
\begin{equation}
 \theta_{i}({\bf{x}}) = {\bf{k}}_{i} \cdot {\bf{x}} 
+ {\tilde{\theta}}_{i}({\bf{x}})
\end{equation}
The reduced system may be written as follows.
\begin{equation}
 \nabla^{2}  {\tilde{\theta}}_{i} 
 + {\bf{k}}_{i} \cdot \nabla {\tilde{\Lambda}}_{i} 
 + \nabla {\tilde{\theta}}_{i} \cdot \nabla {\tilde{\Lambda}}_{i} 
 = 0
\end{equation}
\begin{equation}
\frac{ {\bf{k}}^{2}_{i} }{2m}
 + \frac{ (\nabla {\tilde{\theta}}_{i})^{2} }{2m} 
+ \frac{ {\bf{k}}_{i} \cdot \nabla {\tilde{\theta}}_{i} }{m}
 - \frac{1}{8m} \left[ 2 \nabla^{2} {\tilde{\Lambda}}_{i}
 +  (\nabla {\tilde{\Lambda}}_{i})^{2} \right]
 = \epsilon_{i} - U_{dis}
\end{equation}
Now we make the assertion that $ <{\tilde{\theta}}_{i}> = 0 $
 and $ < {\tilde{\Lambda}}_{i} > = 0 $.
By taking the average of the above equations we may deduce that
\begin{equation}
 \left< \nabla {\tilde{\theta}}_{i} \cdot \nabla {\tilde{\Lambda}}_{i}  \right> = 0
\end{equation}
\begin{equation}
\frac{ {\bf{k}}^{2}_{i} }{2m}
 + \frac{ < (\nabla {\tilde{\theta}}_{i})^{2} > }{2m} 
 - \frac{1}{8m}   \left< (\nabla {\tilde{\Lambda}}_{i})^{2} \right>
 = \epsilon_{i} 
\end{equation}
From this we may also deduce,
\begin{equation}
 \nabla^{2}  {\tilde{\theta}}_{i} 
 + {\bf{k}}_{i} \cdot \nabla {\tilde{\Lambda}}_{i} 
 \approx 0
\end{equation}
\begin{equation}
 \frac{ {\bf{k}}_{i} \cdot \nabla {\tilde{\theta}}_{i} }{m}
 - \frac{ \nabla^{2} {\tilde{\Lambda}}_{i} }{4m}
 \approx - U_{dis}
\end{equation}
 \noindent The above two equations may be used
 to iteratively compute the various correlation functions. First we have,
\begin{equation}
 \nabla^{2}  <{\tilde{\theta}}_{i}({\bf{x}})U_{dis}({\bf{x}}^{'}) > 
 + {\bf{k}}_{i} \cdot \nabla 
<{\tilde{\Lambda}}_{i}({\bf{x}}) U_{dis}({\bf{x}}^{'}) > 
 \approx 0
\end{equation}
\begin{equation}
 \frac{ {\bf{k}}_{i} \cdot \nabla 
<{\tilde{\theta}}_{i}({\bf{x}}) U_{dis}({\bf{x}}^{'}) > }{m}
 - \frac{ \nabla^{2} < {\tilde{\Lambda}}_{i}({\bf{x}}) U_{dis}({\bf{x}}^{'})>  }{4m}
 \approx - <U_{dis}({\bf{x}}) U_{dis}({\bf{x}}^{'}) >
 = - \Delta^{2} \mbox{        }\delta^{d}({\bf{x}}-{\bf{x}}^{'})
\end{equation}
We solve this by Fourier transforms.
\begin{equation}
<{\tilde{\theta}}_{i}({\bf{x}}) U_{dis}({\bf{x}}^{'}) >
 = \frac{1}{V} \sum_{ {\bf{q}} }
F_{ 10 }( {\bf{q}} i )
 \mbox{        } e^{ -i {\bf{q}}.({\bf{x}}-{\bf{x}}^{'}) }
\end{equation}
\begin{equation}
<{\tilde{\Lambda}}_{i}({\bf{x}}) U_{dis}({\bf{x}}^{'}) >
 = \frac{1}{V} \sum_{ {\bf{q}} }
F_{ 20 }( {\bf{q}} i )
\mbox{         } e^{ -i {\bf{q}}.({\bf{x}}-{\bf{x}}^{'}) }
\end{equation}
Thus we have,
\begin{equation}
-{\bf{q}}^{2}\mbox{      }F_{ 10 }( {\bf{q}} i )
- i \mbox{     }({\bf{k}}_{i} \cdot {\bf{q}})
F_{ 20 }( {\bf{q}} i ) = 0
\end{equation}
\begin{equation}
- i \mbox{     }\frac{ {\bf{k}}_{i} \cdot {\bf{q}} }{m}
 F_{ 10 }( {\bf{q}} i )
 + \frac{ {\bf{q}}^{2} }{4m} F_{ 20 }( {\bf{q}} i )
 = - \Delta^{2}
\end{equation}
\begin{equation}
F_{ 10 }( {\bf{q}} i ) = \left( -\frac{ {\bf{q}}^{4} }{4m}
 + \frac{ ({\bf{k}}_{i} \cdot {\bf{q}})^{2} }{m} \right)^{-1}
[-i ({\bf{k}}_{i} \cdot {\bf{q}}) \Delta^{2}]
\end{equation}

\begin{equation}
F_{ 20 }( {\bf{q}} i ) = \left( -\frac{ {\bf{q}}^{4} }{4m}
 + \frac{ ({\bf{k}}_{i} \cdot {\bf{q}})^{2} }{m} \right)^{-1}
[ {\bf{q}}^{2}  \Delta^{2}]
\end{equation}

\begin{equation}
 \nabla^{2}  <{\tilde{\theta}}_{i}({\bf{x}}){\tilde{\Lambda}}_{j}({\bf{x}}^{'}) > 
 + {\bf{k}}_{i} \cdot \nabla 
<{\tilde{\Lambda}}_{i}({\bf{x}}){\tilde{\Lambda}}_{j} ({\bf{x}}^{'}) > 
 \mbox{         }\approx 0
\end{equation}
\begin{equation}
 \frac{ {\bf{k}}_{i} \cdot \nabla 
<{\tilde{\theta}}_{i}({\bf{x}}) {\tilde{\Lambda}}_{j}({\bf{x}}^{'}) > }{m}
 - \frac{ \nabla^{2} < {\tilde{\Lambda}}_{i}({\bf{x}}) {\tilde{\Lambda}}_{j}({\bf{x}}^{'}) >
  }{4m}
 \approx - <U_{dis}({\bf{x}}) {\tilde{\Lambda}}_{j}({\bf{x}}^{'})>
\end{equation}

\begin{equation}
<{\tilde{\theta}}_{i}({\bf{x}}) {\tilde{\Lambda}}_{j}({\bf{x}}^{'}) >
 = \frac{1}{V} \sum_{ {\bf{q}} }
F_{ 12 }( {\bf{q}}; ij )
 \mbox{        } e^{ -i {\bf{q}}.({\bf{x}}-{\bf{x}}^{'}) }
\end{equation}
\begin{equation}
<{\tilde{\Lambda}}_{i}({\bf{x}}){\tilde{\Lambda}}_{j} ({\bf{x}}^{'}) >
 = \frac{1}{V} \sum_{ {\bf{q}} }
F_{ 22 }( {\bf{q}}; ij )
\mbox{         } e^{ -i {\bf{q}}.({\bf{x}}-{\bf{x}}^{'}) }
\end{equation}
\begin{equation}
F_{ 12 }( {\bf{q}}; ij ) = \left( -\frac{ {\bf{q}}^{4} }{4m}
 + \frac{ ({\bf{k}}_{i} \cdot {\bf{q}})^{2} }{m} \right)^{-1}
[-i ({\bf{k}}_{i} \cdot {\bf{q}}) F_{20}(-{\bf{q}}j) ]
\end{equation}
\begin{equation}
F_{ 22 }( {\bf{q}}; ij ) = \left( -\frac{ {\bf{q}}^{4} }{4m}
 + \frac{ ({\bf{k}}_{i} \cdot {\bf{q}})^{2} }{m} \right)^{-1}
[ {\bf{q}}^{2}  F_{20}(-{\bf{q}}j) ]
\end{equation}

\begin{equation}
 \nabla^{2}  <{\tilde{\theta}}_{i}({\bf{x}}){\tilde{ \theta }}_{j}
({\bf{x}}^{'}) > 
 + {\bf{k}}_{i} \cdot \nabla 
<{\tilde{\Lambda}}_{i}({\bf{x}}){\tilde{ \theta }}_{j} ({\bf{x}}^{'}) > 
 \mbox{         }\approx 0
\end{equation}
\begin{equation}
 \frac{ {\bf{k}}_{i} \cdot \nabla 
<{\tilde{\theta}}_{i}({\bf{x}}) {\tilde{ \theta }}_{j}({\bf{x}}^{'}) > }{m}
 - \frac{ \nabla^{2} < {\tilde{\Lambda}}_{i}({\bf{x}}) {\tilde{ \theta }}_{j}
({\bf{x}}^{'}) >
  }{4m}
 \approx - <U_{dis}({\bf{x}}) {\tilde{ \theta }}_{j}({\bf{x}}^{'})>
\end{equation}

\begin{equation}
<{\tilde{\theta}}_{i}({\bf{x}}){\tilde{ \theta }}_{j}({\bf{x}}^{'}) >
 = \frac{1}{V} \sum_{ {\bf{q}} }
F_{ 11 }( {\bf{q}}; ij )
 \mbox{        } e^{ -i {\bf{q}}.({\bf{x}}-{\bf{x}}^{'}) }
\end{equation}
\begin{equation}
<{\tilde{\Lambda}}_{i}({\bf{x}}) {\tilde{ \theta }}_{j} ({\bf{x}}^{'}) >
 = \frac{1}{V} \sum_{ {\bf{q}} }
F_{ 21 }( {\bf{q}}; ij )
\mbox{         } e^{ -i {\bf{q}}.({\bf{x}}-{\bf{x}}^{'}) }
\end{equation}

\begin{equation}
F_{ 11 }( {\bf{q}}; ij ) = \left( -\frac{ {\bf{q}}^{4} }{4m}
 + \frac{ ({\bf{k}}_{i} \cdot {\bf{q}})^{2} }{m} \right)^{-1}
[-i ({\bf{k}}_{i} \cdot {\bf{q}}) F_{10}(-{\bf{q}}j) ]
\end{equation}

\begin{equation}
F_{ 21 }( {\bf{q}}; ij ) = \left( -\frac{ {\bf{q}}^{4} }{4m}
 + \frac{ ({\bf{k}}_{i} \cdot {\bf{q}})^{2} }{m} \right)^{-1}
[ {\bf{q}}^{2}  F_{10}(-{\bf{q}}j) ]
\end{equation}
From the above equations it is clear that,
\[
< \nabla \theta_{i} \cdot \nabla {\tilde{\Lambda}}_{i} >
 = (\nabla_{ {\bf{x}} }
 \cdot \nabla_{ {\bf{x}}^{'} })|_{ {\bf{x}} = {\bf{x}}^{'} } 
\mbox{        }<{\tilde{\theta}}_{i}({\bf{x}}) {\tilde{\Lambda}}_{i}({\bf{x}}^{'})>
\]
\begin{equation}
 = \frac{1}{V} \sum_{ {\bf{q}} }F_{12}({\bf{q}} ; ij)
 \mbox{        }{\bf{q}}^{2} = 0
\end{equation}
\begin{equation}
\epsilon_{i} = \frac{ {\bf{k}}^{2}_{i} }{2m}
 - \frac{ \Delta^{2} }{2V}
\sum_{ {\bf{q}} } {\bf{q}}^{2} 
 \mbox{            }\left( - \frac{ {\bf{q}}^{4} }{4m}
 + \frac{ ({\bf{k}}_{i} \cdot {\bf{q}})^{2} }{m} \right)^{-1}
\end{equation}
In one dimension, we have,
\begin{equation}
\epsilon_{i} = \frac{ k^{2}_{i} }{2m}
 + \frac{ m\Delta^{2} }{ 2\pi k_{i} }
\int^{ \infty }_{0} 
 \mbox{            } dx \mbox{        }
\left( \frac{1}{ x - k_{i} } -  \frac{1}{ x + k_{i} } \right)
\end{equation}
 If we interpret the above integral as the principal part then we have,
\begin{equation}
\epsilon_{i} = \frac{ k^{2}_{i} }{2m}
\end{equation}
 In two space dimensions, it appears that we have to be more careful. In
 particular, we have to introduce a large momentum cutoff that may not
 be easily dropped.
\begin{equation}
\epsilon_{i} = \frac{ k^{2}_{i} }{2m}
 + \frac{ m\Delta^{2} }{ \pi }
\int^{  \frac{ \Lambda }{ 2|k_{i}| }  }_{1} 
\frac{ dx }{ \sqrt{ x^{2} - 1 } }
\end{equation}
 We take the point of view that this may be absorbed by a
 suitable redefinition of the Fermi energy. Thus in all three
 dimensions, we take the liberty
 to set $ \epsilon_{i} = k^{2}_{i}/2m $.

\newpage

\section{Appendix B}

Thus,
\[
\left< {\bf{J}}_{ij}(U_{dis}) \cdot {\bf{J}}_{ji}(U_{dis}) \right>   =
 -\int d^{d}x \int d^{d}y 
\mbox{       }\delta^{d}({\bf{y}}-{\bf{x}})
\int d^{d}x^{'} \int d^{d}y^{'} 
\mbox{       }\delta^{d}({\bf{y}}^{'}-{\bf{x}}^{'})
\]
\[
 ( \nabla_{ {\bf{y}} } \cdot  \nabla_{ {\bf{y}}^{'} })\mbox{        }
\left< \phi^{*}_{i}({\bf{x}})
\phi_{j}({\bf{y}})
\phi^{*}_{j}({\bf{x}}^{'})
\phi_{i}({\bf{y}}^{'}) \right>
\]
\[
  =
 -\frac{1}{V^{2}}
\int d^{d}x \int d^{d}y 
\mbox{       }\delta^{d}({\bf{y}}-{\bf{x}})
\int d^{d}x^{'} \int d^{d}y^{'} 
\mbox{       }\delta^{d}({\bf{y}}^{'}-{\bf{x}}^{'})
\]
\[
 ( \nabla_{ {\bf{y}} } \cdot  \nabla_{ {\bf{y}}^{'} })\mbox{        }
e^{i{\bf{k}}_{i} \cdot ({\bf{y}}^{'}-{\bf{x}}) }
e^{i{\bf{k}}_{j} \cdot ({\bf{y}}-{\bf{x}}^{'}) }
\left< e^{i {\tilde{\theta}}_{j}({\bf{y}}) + \frac{1}{2}{\tilde{\Lambda}}_{j}({\bf{y}}) }
e^{i {\tilde{\theta}}_{i}({\bf{y}}^{'}) + \frac{1}{2}
{\tilde{\Lambda}}_{i}({\bf{y}}^{'}) }
e^{-i {\tilde{\theta}}_{i}({\bf{x}}) + \frac{1}{2}{\tilde{\Lambda}}_{i}({\bf{x}}) }
e^{-i {\tilde{\theta}}_{j}({\bf{x}}^{'}) + \frac{1}{2}
{\tilde{\Lambda}}_{j}({\bf{x}}^{'}) }
\right>
\]
Thus we have,
\[
\left< e^{i {\tilde{\theta}}_{j}({\bf{y}}) + \frac{1}{2}
{\tilde{\Lambda}}_{j}({\bf{y}}) }
e^{i {\tilde{\theta}}_{i}({\bf{y}}^{'}) + \frac{1}{2}
{\tilde{\Lambda}}_{i}({\bf{y}}^{'}) }
e^{-i {\tilde{\theta}}_{i}({\bf{x}}) + \frac{1}{2}
{\tilde{\Lambda}}_{i}({\bf{x}}) }
e^{-i {\tilde{\theta}}_{j}({\bf{x}}^{'}) + \frac{1}{2}
{\tilde{\Lambda}}_{j}({\bf{x}}^{'}) }
\right>
 = e^{\frac{1}{2}<E^{2}> }
\]
\[
<E^{2}> = 
 \left< ( i {\tilde{\theta}}_{j}({\bf{y}}) + \frac{1}{2}{\tilde{\Lambda}}_{j}({\bf{y}}) 
+ i {\tilde{\theta}}_{i}({\bf{y}}^{'}) + \frac{1}{2}
{\tilde{\Lambda}}_{i}({\bf{y}}^{'}) 
-i {\tilde{\theta}}_{i}({\bf{x}}) + \frac{1}{2}{\tilde{\Lambda}}_{i}({\bf{x}}) 
-i {\tilde{\theta}}_{j}({\bf{x}}^{'}) + \frac{1}{2}
{\tilde{\Lambda}}_{j}({\bf{x}}^{'}))^{2} \right>
\]
\[
 = -<  {\tilde{\theta}}^{2}_{j}({\bf{y}}) >
 + \frac{1}{4}< {\tilde{\Lambda}}^{2}_{j}({\bf{y}}) >
- < {\tilde{\theta}}^{2}_{i}({\bf{y}}^{'}) >
 +  \frac{1}{4}  < {\tilde{\Lambda}}^{2}_{i}({\bf{y}}^{'}) >
\]
\[
- < {\tilde{\theta}}^{2}_{i}({\bf{x}}) >
+ \frac{1}{4}  < {\tilde{\Lambda}}^{2}_{i}({\bf{x}}) >
- < {\tilde{\theta}}^{2}_{j}({\bf{x}}^{'}) >
+  \frac{1}{4}  < {\tilde{\Lambda}}^{2}_{j}({\bf{x}}^{'}) >
\]
\[
+ i \mbox{   }<  {\tilde{\theta}}_{j}({\bf{y}}) {\tilde{\Lambda}}_{j}({\bf{y}}) >
-2 <{\tilde{\theta}}_{j}({\bf{y}}) 
{\tilde{\theta}}_{i}({\bf{y}}^{'}) >
+ i < {\tilde{\theta}}_{j}({\bf{y}}) {\tilde{\Lambda}}_{i}({\bf{y}}^{'}) >
+ 2 < {\tilde{\theta}}_{j}({\bf{y}})  {\tilde{\theta}}_{i}({\bf{x}}) >
+ i <{\tilde{\theta}}_{j}({\bf{y}})  {\tilde{\Lambda}}_{i}({\bf{x}}) >
\]
\[
+ 2 <{\tilde{\theta}}_{j}({\bf{y}})  {\tilde{\theta}}_{j}({\bf{x}}^{'}) >
+ i <  {\tilde{\theta}}_{j}({\bf{y}})  {\tilde{\Lambda}}_{j}({\bf{x}}^{'}) >
+ i < {\tilde{\Lambda}}_{j}({\bf{y}}) {\tilde{\theta}}_{i}({\bf{y}}^{'}) >
+ \frac{1}{2} < {\tilde{\Lambda}}_{j}({\bf{y}}) {\tilde{\Lambda}}_{i}({\bf{y}}^{'}) >
\]
\[
-i < {\tilde{\Lambda}}_{j}({\bf{y}})  {\tilde{\theta}}_{i}({\bf{x}}) >
+ \frac{1}{2} < {\tilde{\Lambda}}_{j}({\bf{y}})  {\tilde{\Lambda}}_{i}({\bf{x}}) >
-i < {\tilde{\Lambda}}_{j}({\bf{y}}) {\tilde{\theta}}_{j}({\bf{x}}^{'}) >
+\frac{1}{2}  < {\tilde{\Lambda}}_{j}({\bf{y}}) {\tilde{\Lambda}}_{j}({\bf{x}}^{'})  >
\]
\[
+ i < {\tilde{\theta}}_{i}({\bf{y}}^{'}) {\tilde{\Lambda}}_{i}({\bf{y}}^{'}) >
+ 2  < {\tilde{\theta}}_{i}({\bf{y}}^{'})  {\tilde{\theta}}_{i}({\bf{x}}) >
+ i < {\tilde{\theta}}_{i}({\bf{y}}^{'})  {\tilde{\Lambda}}_{i}({\bf{x}}) >
+ 2  < {\tilde{\theta}}_{i}({\bf{y}}^{'}) {\tilde{\theta}}_{j}({\bf{x}}^{'}) >
+ i < {\tilde{\theta}}_{i}({\bf{y}}^{'}) {\tilde{\Lambda}}_{j}({\bf{x}}^{'}) >
\]
\[
-i <{\tilde{\Lambda}}_{i}({\bf{y}}^{'})   {\tilde{\theta}}_{i}({\bf{x}}) >
+ \frac{1}{2} <  {\tilde{\Lambda}}_{i}({\bf{y}}^{'})   {\tilde{\Lambda}}_{i}({\bf{x}}) >
- i < {\tilde{\Lambda}}_{i}({\bf{y}}^{'}) {\tilde{\theta}}_{j}({\bf{x}}^{'}) >
 + \frac{1}{2} < {\tilde{\Lambda}}_{i}({\bf{y}}^{'})  {\tilde{\Lambda}}_{j}({\bf{x}}^{'}) >
\]
\[
- i < {\tilde{\theta}}_{i}({\bf{x}}) {\tilde{\Lambda}}_{i}({\bf{x}})  >
-2 < {\tilde{\theta}}_{i}({\bf{x}}) {\tilde{\theta}}_{j}({\bf{x}}^{'}) >
- i < {\tilde{\theta}}_{i}({\bf{x}}) {\tilde{\Lambda}}_{j}({\bf{x}}^{'}) >
\]
\[
- i < {\tilde{\Lambda}}_{i}({\bf{x}}) {\tilde{\theta}}_{j}({\bf{x}}^{'}) >
+ \frac{1}{2} < {\tilde{\Lambda}}_{i}({\bf{x}}) {\tilde{\Lambda}}_{j}({\bf{x}}^{'}) >
- i <{\tilde{\theta}}_{j}({\bf{x}}^{'}) {\tilde{\Lambda}}_{j}({\bf{x}}^{'})>
\]

\[
<E^{2}> = E_{0}(i,j) + \frac{1}{V} \sum_{ {\bf{q}} }
A_{22^{'}}({\bf{q}};i,j) \mbox{         }
e^{-i{\bf{q}}.({\bf{y}}-{\bf{y}}^{'})}
+ \frac{1}{V} \sum_{ {\bf{q}} }
A_{2^{'}1}({\bf{q}};i,j) \mbox{         }
e^{-i{\bf{q}}.({\bf{y}}^{'}-{\bf{x}})}
\]
\[
 + \frac{1}{V} \sum_{ {\bf{q}} }
A_{21^{'}}({\bf{q}};i,j) \mbox{         }
e^{-i{\bf{q}}.({\bf{y}}-{\bf{x}}^{'})}
+ \frac{1}{V} \sum_{ {\bf{q}} }
A_{2^{'}1^{'}}({\bf{q}};i,j) \mbox{         }
e^{-i{\bf{q}}.({\bf{y}}^{'}-{\bf{x}}^{'})}
\]
\begin{equation}
 + \frac{1}{V} \sum_{ {\bf{q}} }
A_{21}({\bf{q}};i,j) \mbox{         }
e^{-i{\bf{q}}.({\bf{y}}-{\bf{x}})}
+ \frac{1}{V} \sum_{ {\bf{q}} }
A_{11^{'}}({\bf{q}};i,j) \mbox{         }
e^{-i{\bf{q}}.({\bf{x}}-{\bf{x}}^{'})}
\end{equation}
\begin{equation}
 E_{0}(i,j) = -\frac{2}{V} \sum_{ {\bf{q}} }F_{11}({\bf{q}};jj)
 + \frac{1}{2V} \sum_{ {\bf{q}} }F_{22}({\bf{q}};jj)
- \frac{2}{V} \sum_{ {\bf{q}} } F_{11}({\bf{q}};ii)
+ \frac{1}{2V} \sum_{ {\bf{q}} }F_{22}({\bf{q}};ii)
\end{equation}
\begin{equation}
A_{22^{'}}({\bf{q}};i,j) = -2 \mbox{   }F_{11}({\bf{q}};ji)
 + i \mbox{   }F_{12}({\bf{q}};ji)
 + i\mbox{       }F_{21}({\bf{q}};ji) + \frac{1}{2}\mbox{    }
 F_{22}({\bf{q}};ji)
\end{equation}
\begin{equation}
A_{2^{'}1}({\bf{q}};i,j) = 2\mbox{       }
 F_{11}({\bf{q}};ii)
 + i \mbox{       } F_{12}({\bf{q}};ii)
 - i \mbox{       }
F_{21}({\bf{q}};ii) + \frac{1}{2} \mbox{         }
 F_{22}({\bf{q}};ii)
\end{equation}
\begin{equation}
A_{21^{'}}({\bf{q}};i,j) = 2 \mbox{        }
 F_{11}({\bf{q}};jj)
 + i \mbox{        }
 F_{12}({\bf{q}};jj)
 - i \mbox{        }
F_{21}({\bf{q}};jj) + \frac{1}{2} \mbox{     } F_{22}({\bf{q}};jj)
\end{equation}
\begin{equation}
A_{2^{'}1^{'}}({\bf{q}};i,j) = 2 \mbox{       }
 F_{11}({\bf{q}};ij)
 + i \mbox{         }
 F_{12}({\bf{q}};ij)
 - i \mbox{        }
F_{21}({\bf{q}};ij) + \frac{1}{2} \mbox{      } F_{22}({\bf{q}};ij)
\end{equation}
\begin{equation}
A_{21}({\bf{q}};i,j) = 2 \mbox{     } F_{11}({\bf{q}};ji)
 + i \mbox{        }F_{12}({\bf{q}};ji)
 - i \mbox{       }
F_{21}({\bf{q}};ji) + \frac{1}{2} \mbox{       } F_{22}({\bf{q}};ji)
\end{equation}
\begin{equation}
A_{11^{'}}({\bf{q}};i,j) = -2 \mbox{   }
 F_{11}({\bf{q}};ij)
 - i \mbox{       } F_{12}({\bf{q}};ij)
 - i \mbox{       }
F_{21}({\bf{q}};ij) + \frac{1}{2} \mbox{      }
 F_{22}({\bf{q}};ij)
\end{equation}
\begin{equation}
\left< {\bf{J}}_{ij}(U_{dis}) \cdot {\bf{J}}_{ji}(U_{dis}) \right>   =
 -\frac{ C^{0}_{i} \mbox{     }C^{0}_{j} }{V^{2}}
\int d^{d}x  
\mbox{       }
\int d^{d}x^{'} \mbox{       }
e^{i({\bf{k}}_{j} - {\bf{k}}_{i})\cdot ({\bf{x}}-{\bf{x}}^{'}) }
e^{\frac{1}{2}<E^{2}> }
F({\bf{k}}_{i},{\bf{k}}_{j})
\label{JIJSQ}
\end{equation}
\[
F({\bf{k}}_{i},{\bf{k}}_{j})
 = \frac{1}{2V} \sum_{ {\bf{q}} }A_{22^{'}}({\bf{q}};ij)
 \mbox{   }e^{-i{\bf{q}}.({\bf{x}}-{\bf{x}}^{'})} \mbox{       }{\bf{q}}^{2}
\]
\[
+ [ i{\bf{k}}_{i} 
 + \frac{1}{2V}
\sum_{ {\bf{q}} }A_{2^{'}1^{'}}({\bf{q}};ij) (-i{\bf{q}})
+ \frac{1}{2V}
\sum_{ {\bf{q}} }\left( A_{22^{'}}({\bf{q}};ij) (i{\bf{q}})
 + A_{2^{'}1}(-{\bf{q}};ij) (i{\bf{q}}) \right)
e^{-i{\bf{q}}.({\bf{x}}-{\bf{x}}^{'})} ]
\]
\begin{equation}
\times \mbox{     }
 [ i{\bf{k}}_{j} + \frac{1}{2V}
\sum_{ {\bf{q}} }A_{21}({\bf{q}};ij) (-i{\bf{q}})
+ \frac{1}{2V}
\sum_{ {\bf{q}} }\left( A_{22^{'}}({\bf{q}};ij) (-i{\bf{q}})
 + A_{21^{'}}({\bf{q}};ij) (-i{\bf{q}}) \right)
e^{-i{\bf{q}}.({\bf{x}}-{\bf{x}}^{'})} ]
\end{equation}
\[
\frac{1}{2} <E^{2}> = -\frac{1}{V}\sum_{ {\bf{q}} }F_{11}({\bf{q}};jj)
 + \frac{1}{4V}\sum_{ {\bf{q}} }F_{22}({\bf{q}};jj)
 - \frac{1}{V}\sum_{ {\bf{q}} }F_{11}({\bf{q}};ii)
\]
\[
+ \frac{1}{4V}\sum_{ {\bf{q}} }F_{22}({\bf{q}};ii)
+ \frac{1}{2V} \sum_{ {\bf{q}} }A_{21}({\bf{q}};i,j)
+ \frac{1}{2V} \sum_{ {\bf{q}} }A_{2^{'}1^{'}}({\bf{q}};i,j)
\]
\begin{equation}
+ \frac{1}{2V} \sum_{ {\bf{q}} }
\left[ A_{22^{'}}({\bf{q}};i,j)
 + A_{2^{'}1}(-{\bf{q}};i,j)
 + A_{21^{'}}({\bf{q}};i,j)
 + A_{11^{'}}({\bf{q}};i,j) \right] \mbox{       }
e^{-i{\bf{q}}.({\bf{x}}-{\bf{x}}^{'})}
\end{equation}
\begin{equation}
F_{12}({\bf{q}};ij) = P_{i}({\bf{q}})P_{j}({\bf{q}})
[-i({\bf{k}}_{i}.{\bf{q}}){\bf{q}}^{2} \Delta^{2}]
\end{equation}
\begin{equation}
F_{22}({\bf{q}};ij) =
P_{i}({\bf{q}})P_{j}({\bf{q}})[{\bf{q}}^{4} \Delta^{2}]
\end{equation}
\begin{equation}
F_{11}({\bf{q}};ij) = 
P_{i}({\bf{q}})P_{j}({\bf{q}})[ ({\bf{k}}_{i}.{\bf{q}})  ({\bf{k}}_{j}.{\bf{q}})  \Delta^{2}]
\end{equation}
\begin{equation}
F_{21}({\bf{q}};ij) = 
P_{i}({\bf{q}})P_{j}({\bf{q}})
[ i({\bf{k}}_{j}.{\bf{q}})  {\bf{q}}^{2}  \Delta^{2}]
\end{equation}
\begin{equation}
 P_{i}({\bf{q}}) = \left( -\frac{ {\bf{q}}^{4} }{4m}
 + \frac{ ({\bf{k}}_{i}.{\bf{q}})^{2} }{m} \right)^{-1}
\end{equation}
Define,
\begin{equation}
D_{i}({\bf{q}}) = -{\bf{k}}_{i}.{\bf{q}} + \frac{ {\bf{q}}^{2} }{2}
\end{equation}
\begin{equation}
A_{22^{'}}({\bf{q}};i,j) 
 =  A_{11^{'}}({\bf{q}};i,j) =
\frac{ 2 m^{2} \Delta^{2}  }
{ D_{i}(-{\bf{q}})D_{j}({\bf{q}}) }
\end{equation}
\begin{equation}
A_{2^{'}1}({\bf{q}};i,j)
 =  \frac{ 2 m^{2} \Delta^{2} }{ D^{2}_{i}({\bf{q}}) }
\end{equation}
\begin{equation}
A_{21^{'}}({\bf{q}};i,j)
 =  \frac{ 2 m^{2} \Delta^{2} }{ D^{2}_{j}({\bf{q}}) }
\end{equation}
\begin{equation}
A_{2^{'}1^{'}}({\bf{q}};i,j)
= A_{21}({\bf{q}};i,j)
=  \frac{ 2m^{2} \Delta^{2} }{ D_{i}({\bf{q}})
D_{j}({\bf{q}}) }
\end{equation}
\begin{equation}
C^{0}_{i} = exp \left[ - \frac{ m^{2} \Delta^{2} }{2V} \sum_{ {\bf{q}} }
\left( \frac{1}{ D_{i}({\bf{q}}) } +  \frac{1}{ D_{i}(-{\bf{q}}) } \right)^{2}
\right]
\end{equation}
Define,
\begin{equation}
 \frac{1}{2} <E^{'2}> =  \frac{1}{2} <E^{2}>
 - \frac{ m^{2} \Delta^{2} }{2V} \sum_{ {\bf{q}} }
\left( \frac{1}{ D_{i}({\bf{q}}) } + \frac{1}{ D_{i}(-{\bf{q}}) } \right)^{2}
 - \frac{ m^{2} \Delta^{2} }{2V} \sum_{ {\bf{q}} }
\left( \frac{1}{ D_{j}({\bf{q}}) } + \frac{1}{ D_{j}(-{\bf{q}}) } \right)^{2}
\end{equation}
\begin{equation}
\frac{1}{2} <E^{'2}> =
 - \frac{ m^{2} \Delta^{2} }{V} \sum_{ {\bf{q}} }
 \left( \frac{1}{ D_{i}({\bf{q}}) }  - \frac{1}{ D_{j}({\bf{q}}) } \right)^{2} 
+ \frac{ m^{2} \Delta^{2} }{V} \sum_{ {\bf{q}} }
\left( \frac{ 1 }{ D_{i}(-{\bf{q}}) }
 + \frac{ 1 }{ D_{j}({\bf{q}}) } \right)^{2}
e^{-i{\bf{q}}.({\bf{x}}-{\bf{x}}^{'})}
\end{equation}
The above equation leads to a finite result as we may see below.
\[
\frac{1}{2} <E^{'2}> =
 \frac{ m^{2} \Delta^{2} }{V} \sum_{ {\bf{q}} }
\left( \frac{1}{D_{j}(-{\bf{q}})}
+  \frac{1}{D_{j}({\bf{q}})} \right)
 \mbox{        } \left( \frac{1}{ D_{i}(-{\bf{q}}) } 
 + \frac{1}{ D_{i}({\bf{q}}) } \right)
\]
\[
-  \mbox{   }\frac{ m^{2} \Delta^{2} }{V} \sum_{ {\bf{q}} }
\left( \frac{ 1 }{ D_{i}(-{\bf{q}}) }
 + \frac{ 1 }{ D_{j}({\bf{q}}) } \right)^{2} 
\mbox{       }\left( 1 - cos[{\bf{q}}.({\bf{x}} - {\bf{x}}^{'})]  \right)
\]
\[
- i \mbox{   }\frac{ m^{2} \Delta^{2} }{2V} \sum_{ {\bf{q}} }
 \left( \frac{ 1 }{ D_{i}(-{\bf{q}}) }
  +  \frac{ 1 }{ D_{i}({\bf{q}}) }
 + \frac{ 1 }{ D_{j}({\bf{q}}) }  
 + \frac{ 1 }{ D_{j}(-{\bf{q}}) } \right)
\]
\begin{equation}
 \times \left( \frac{ 1 }{ D_{i}(-{\bf{q}}) }
 + \frac{ 1 }{ D_{j}({\bf{q}}) }  
  -  \frac{ 1 }{ D_{i}({\bf{q}}) }
 - \frac{ 1 }{ D_{j}(-{\bf{q}}) } \right)
sin[{\bf{q}}.({\bf{x}} - {\bf{x}}^{'})]
\end{equation}
Similarly, we may write for $ F({\bf{k}}_{i},{\bf{k}}_{j}) $ as follows,
\[
F({\bf{k}}_{i},{\bf{k}}_{j})
 = \frac{1}{2V} \sum_{ {\bf{q}} }\frac{2 m^{2} \Delta^{2} {\bf{q}}^{2} }
{ D_{i}(-{\bf{q}})D_{j}({\bf{q}})} 
e^{-i{\bf{q}}.({\bf{x}}-{\bf{x}}^{'})}
\]
\[
+ [ i{\bf{k}}_{i} 
 + \frac{1}{2V}
\sum_{ {\bf{q}} }
\frac{ 2 m^{2} \Delta^{2} }{ D_{i}({\bf{q}})D_{j}({\bf{q}}) }
 (-i{\bf{q}})
+ \frac{1}{2V}
\sum_{ {\bf{q}} }\left( 
\frac{ 2 m^{2} \Delta^{2} }{ D_{i}(-{\bf{q}})D_{j}({\bf{q}}) }
 (i{\bf{q}})
 + \frac{ 2 m^{2} \Delta^{2} }{ D^{2}_{i}(-{\bf{q}}) }
 (i{\bf{q}}) \right)
e^{-i{\bf{q}}.({\bf{x}}-{\bf{x}}^{'})} ]
\]
\begin{equation}
\cdot \mbox{     }
 [ i{\bf{k}}_{j} + \frac{1}{2V}
\sum_{ {\bf{q}} }
\frac{ 2 m^{2} \Delta^{2} }{ D_{i}({\bf{q}})D_{j}({\bf{q}}) }
 (-i{\bf{q}})
+ \frac{1}{2V}
\sum_{ {\bf{q}} }\left( 
\frac{ 2 m^{2} \Delta^{2} }{ D_{i}(-{\bf{q}})D_{j}({\bf{q}}) }
 (-i{\bf{q}})
 +  \frac{ 2 m^{2} \Delta^{2} }{ D^{2}_{j}({\bf{q}}) }
 (-i{\bf{q}}) \right)
e^{-i{\bf{q}}.({\bf{x}}-{\bf{x}}^{'})} ]
\end{equation}
 Now we would like to solve for $ F({\bf{k}}_{i},{\bf{k}}_{j}) $ and
 $ \frac{1}{2}<E^{'2}> $ in
 one dimension and three dimensions where the integrals are likely to
 be simple. 

\vspace{0.2in}

\noindent {\bf{B.1}} : {\bf{ One Dimension }} :

\begin{equation}
\frac{1}{2} < E^{'2} > = - \frac{ m^{2} \Delta^{2} }{ 2 }
\left( \frac{1}{ k_{i} } - \frac{1}{ k_{j} } \right)^{2} 
\mbox{            } |x| 
+ i \frac{ m^{2} \Delta^{2} }{ k^{2}_{F} }
sgn(x)
\left( \frac{1}{ k_{i} } - \frac{1}{ k_{j} } \right)
\end{equation}
\[
F(k_{i},k_{j}) = -i \frac{ m^{2} \Delta^{2} }{ k_{i} + k_{j} }
\left( e^{ -2i k_{j}x } 
 - e^{ 2i k_{i}x } \right) \mbox{        }sgn(x)
\]
\[
+ [ ik_{i} 
- \frac{ m^{2} \Delta^{2} }{2} 
\{ \frac{1}{ k_{i}k_{j} }
 - \frac{ e^{ -2ik_{j}x } }{ k_{j}(k_{j}+k_{i}) }
 - \frac{ e^{ 2ik_{i}x } }{ k_{i}(k_{j}+k_{i}) } 
 +  \frac{1}{ -k_{i}k_{i} }
 + \frac{ e^{ 2ik_{i}x } }{ k_{i} k_{i} } \} sgn(x) ]
\]
\begin{equation}
\times \mbox{     }
 [ ik_{j} 
 + \frac{ m^{2} \Delta^{2} }{2} 
\{ \frac{1}{ k_{i}k_{j} }
 - \frac{ e^{ -2ik_{j}x } }{ k_{j}(k_{j}+k_{i}) }
 - \frac{ e^{ 2ik_{i}x } }{ k_{i}(k_{j}+k_{i}) } 
 + \frac{1}{ -k_{j}k_{j} }
 + \frac{ e^{ -2ik_{j}x } }{ k_{j} k_{j} }  \} sgn(x) ]
\end{equation}
\begin{equation}
f(k_{i},k_{j}) \equiv \left< {\bf{J}}_{ij}(U_{dis}) \cdot {\bf{J}}_{ji}(U_{dis}) \right>   =
 -\frac{1}{L}
\int_{-\infty}^{\infty}  dx  
\mbox{       }e^{i(k_{j} - k_{i}) x }e^{\frac{1}{2}<E^{'2}> }
 F(k_{i},k_{j})
\end{equation}
 The final formula for the a.c. conductivity
 involves evaluating the following integral.
  For $ m \omega << k^{2}_{F} $ we have,
\[
Re[\sigma(\omega)] \sim \frac{1}{ \omega }
\left( \int^{ - k_{F} + \omega/v_{F} }_{ -k_{F} } dk_{j}
 +   \int^{ k_{F} }_{ k_{F}-\omega/v_{F} } dk_{j} \right)\mbox{     }
[f(k_{j}(1+m\omega/k^{2}_{F}),k_{j}) 
+ f(-k_{j}(1+m\omega/k^{2}_{F}),k_{j})]
\]
\[
\approx 
 \frac{ 2 }{ v_{F} } 
[ f(k_{F}(1+m\omega/k^{2}_{F}),k_{F}) 
+ f(-k_{F}(1+m\omega/k^{2}_{F}),k_{F})
\]
\begin{equation}
 + f(-k_{F}(1+m\omega/k^{2}_{F}),-k_{F}) 
+ f(k_{F}(1+m\omega/k^{2}_{F}),-k_{F}) ]
\end{equation}
 We would like to systematically evaluate this and show that in the zero
 frequency limit, the a.c. conductivity is proprotional to $ \omega $
 and vanishes for $ \omega = 0 $. The $ \omega \rightarrow 0 $ limit is quite
 subtle. In particular, we may not set $ \omega = 0 $ at the outset.
 For then, $ f(k_{F},k_{F}) = \infty $. 
\begin{equation}
f(k_{F}(1+m\omega/k^{2}_{F}),k_{F})
 \approx  -\frac{1}{L}
\int_{-\infty}^{\infty}  dx  
\mbox{       }e^{ -i\frac{ m \omega }{k_{F}} x }
\mbox{       }
e^{ -i\frac{ m^{3} \Delta^{2} \omega }{ k^{5}_{F} }
  sgn(x) }   F(k_{F},k_{F})
\end{equation}
\[
  F(k_{F},k_{F}) = -i \frac{ m^{2} \Delta^{2} }{2k_{F}}
(e^{-2ik_{F}x} - e^{2ik_{F}x}) \mbox{    }sgn(x)
\]
\begin{equation}
+ [ i k_{F} + \frac{ m^{2} \Delta^{2} }{2}
 \{ - \frac{ e^{ 2 i  k_{F} x }  }
{ 2k^{2}_{F} }
 + \frac{ e^{ -2ik_{F}x } }{ 2k^{2}_{F} } \}sgn(x) ]^{2}
\end{equation}
and,
\[
f(-k_{F}(1+m\omega/k^{2}_{F}),k_{F})
 \approx 
 f(-k_{F},k_{F}) \approx
\]
\begin{equation}
 \mbox{      } \approx \mbox{      }
 -\frac{1}{L}
\int_{-\infty}^{\infty}  dx  
\mbox{       }e^{ 2 i k_{F} x  }
\mbox{       }
e^{ - \frac{2 m^{2} \Delta^{2} }{k^{2}_{F}} |x| }
e^{ -i \frac{ 2 m^{2} \Delta^{2} }{ k^{3}_{F} }  sgn(x)  }
 F(-k_{F},k_{F})
\end{equation}
\[
 F(-k_{F},k_{F}) = - \mbox{     } ( 2 m^{2} \Delta^{2} ) 
\mbox{         }   e^{ -2 i k_{F} x } \mbox{         } x
\]
\begin{equation}
- [ ik_{F} + \frac{ m^{2} \Delta^{2} }{ k^{2}_{F} } 
( -1 +  e^{ -2ik_{F} x}  )
 \mbox{        }sgn(x) ]^{2}
\end{equation}

\begin{equation}
f(-k_{F}(1+m\omega/k^{2}_{F}),-k_{F})
 \approx  -\frac{1}{L}
\int_{-\infty}^{\infty}  dx  
\mbox{       }e^{ i\frac{ m \omega }{k_{F}} x }
\mbox{       }
e^{ i\frac{ m^{3} \Delta^{2} \omega }{ k^{5}_{F} }
  sgn(x) }   F(-k_{F},-k_{F})
\end{equation}
\begin{equation}
  F(-k_{F},-k_{F}) = F(k_{F},k_{F})
\end{equation}
and,
\[
f(k_{F}(1+m\omega/k^{2}_{F}),-k_{F})
 \approx 
 f(k_{F},-k_{F}) \approx
\]
\begin{equation}
 \mbox{      } \approx \mbox{      }
 -\frac{1}{L}
\int_{-\infty}^{\infty}  dx  
\mbox{       }e^{ -2 i k_{F} x  }
\mbox{       }
e^{ - \frac{2 m^{2} \Delta^{2} }{k^{2}_{F}} |x| }
e^{ i \frac{ 2 m^{2} \Delta^{2} }{ k^{3}_{F} }  sgn(x)  }
 F(k_{F},-k_{F})
\end{equation}
\[
 F(k_{F},-k_{F}) = - \mbox{     } ( 2 m^{2} \Delta^{2} ) 
\mbox{         }   e^{ 2 i k_{F} x } \mbox{         } x
\]
\begin{equation}
- [ -ik_{F} + \frac{ m^{2} \Delta^{2} }{ k^{2}_{F} } 
( -1 +  e^{ 2ik_{F} x}  )
 \mbox{        }sgn(x) ]^{2}
\end{equation}

\newpage

\[
f(k_{F}(1+m \omega/k_{F}^{2}),k_{F}) 
 + f(-k_{F}(1+m \omega/k_{F}^{2}),-k_{F}) 
\]
\begin{equation}
 = -\frac{2}{L} 
\frac{ \left[ \left( \Delta^{4} k_{F} m^{5} \omega
  \frac{ m\omega}{ k_{F} }  
+ (8 k^{6}_{F} + \Delta^{4} m^{4}) 
(16 k^{4}_{F} - m^{2} \omega^{2}) \right)
sin[\Delta^{2} m^{3} \omega/k^{5}_{F}] \right] }
{ 4 k^{4}_{F} \frac{ m \omega }{ k_{F} } ( 16 k^{4}_{F} - m^{2} \omega^{2} ) }
\end{equation}
\[
f(k_{F}(1+m \omega/k_{F}^{2}),k_{F}) 
 + f(-k_{F}(1+m \omega/k_{F}^{2}),-k_{F}) 
\]
\begin{equation}
 \approx - \frac{ 8 \Delta^{2} k^{6}_{F} m^{2} + \Delta^{6} m^{6} }
{ 2 k^{8}_{F} L } 
 + \frac{ \Delta^{6} m^{8} (61 k^{6}_{F} + 8 \Delta^{4} m^{4}) \omega^{2} }
{ 96 k^{18}_{F} L }  + {\large{O}}  [\omega^{4}] 
\end{equation}
\begin{equation}
f(-k_{F}(1+m \omega/k_{F}^{2}),k_{F}) 
 = -\frac{ k^{4}_{F} }{L} \frac{ 
\left[ \Delta^{4} m^{4} cos[ 2 \Delta^{2} m^{2}/ k^{3}_{F} ]
+ i ( k^{6}_{F} + i \Delta^{2} k^{3}_{F} m^{2} + \Delta^{4} m^{4} )
sin[2 \Delta^{2} m^{2}/k^{3}_{F}] \right] }{ \Delta^{2} m^{2}
 (k^{6}_{F} + \Delta^{4} m^{4}) }
\end{equation}
 These have been evaluated using $ Mathematica^{TM} $.
 Immediately we see several
 problems. $ f(k_{i},k_{j}) \geq 0 $ for all arguments.
 Yet the above equations tell us that they are negative
 sometimes and sometimes even complex ! 

 This means that something is seriously wrong with our formalism. Perhaps
 replacing $ \nabla {\tilde{\theta}}_{i} \cdot \nabla {\tilde{\Lambda}}_{i} $
 by the average (which is zero !) was not such a good idea after all.

 What is as bad is that the final formula for the a.c. conductivity
 has a nonvanishing constant part (at least it is real !) 
 this means that d.c. conductivity is
 not zero. This is clearly wrong. Maybe some readers of this preprint
 will offer to collaborate with the authors to fix this difficulty.

\vspace{0.2in}

\noindent {\bf{B.2}} : {\bf{Two and Three Dimensions}} :

\vspace{0.2in}

 \noindent In two and three dimensions, the integrals are
 substantially more complicated caused by the complicated angular parts. 
 Work is in progress in collaboration with Shri. Chandradew Sharma
 it will be reported soon. Useful conversations with Prof. N.D. Haridass
 of IMSc. is gratefully acknowledged. 

\section{Appendix C}

 We would like to ascertain whether or not the zero frequency limit
 of the a.c. conductivity is the d.c. conductivity.
 First let us define d.c. conductivity. Consider a system of electrons
 coupled first to a uniform d.c. electric field.
 The interaction part of the hamiltonian may be written as,
\begin{equation}
H_{I} = |e| \int d^{d}x \mbox{   }\psi^{\dagger}({\bf{x}}) 
 \mbox{       }( {\bf{E}}_{ext} \cdot {\bf{x}} ) \mbox{      }
\psi({\bf{x}}) 
\end{equation}
 The expectation value of the total momentum of the electrons may be
 written in the interaction representation as,
\begin{equation}
\left< {\bf{P}}(t) \right> = \frac{ \left< T \mbox{        }S
 \mbox{        } {\bf{ {\hat{P}} }}(t) \right> }
{  \left< T \mbox{        }S  \right> }
\end{equation}
\begin{equation}
S = e^{ -i \int^{ -i \beta }_{0} dt \mbox{        }{\hat{H}}_{I}(t) }
\end{equation}
The d.c. conductivity is then simply given by,
\begin{equation}
\sigma_{d.c.} = \frac{ |e| }{mV}
\left(  \frac{ \delta }{ \delta  {\bf{E}}_{ext} } \left< {\bf{P}}(t) \right>
 \right)_{ {\bf{E}}_{ext} \equiv 0 }
\end{equation}
Since,
\begin{equation}
\frac{ \delta }{ \delta  {\bf{E}}_{ext} }  S
 = -i \int^{ -i \beta }_{0} dt^{'} 
|e| \int d^{d}x^{'} \psi^{\dagger}({\bf{x}}^{'},t^{'}) 
\mbox{  }{\bf{x}}^{'} \mbox{  }
\psi({\bf{x}}^{'},t^{'}) 
\end{equation}
Define,
\begin{equation}
{\bf{ {\hat{X}} }}(t^{'}) = \int d^{d}x^{'} \psi^{\dagger}({\bf{x}}^{'},t^{'}) 
\mbox{  }{\bf{x}}^{'} \mbox{  }
\psi({\bf{x}}^{'},t^{'}) 
\end{equation}
\begin{equation}
\sigma_{d.c.} = -i\mbox{        }\frac{ e^{2} }{mV}
  \int^{ -i \beta }_{0} dt^{'} 
 \left< T \mbox{       }
{\bf{ {\hat{X}} }}(t^{'}) \cdot {\bf{ {\hat{P}} }}(t) \right>
\label{SIGDC}
\end{equation}
 From the main text we see that the zero frequency limit
 of the a.c. conductivity is written as,
\[
\sigma_{a.c.}(0) \equiv
\int^{-i\beta}_{0} dt^{'} \mbox{       }{\tilde{\sigma}}(t-t^{'})
\]
\begin{equation}
 = -\frac{ i \mbox{  }e^{2} }{m^{2}V}
 \mbox{         }
\int^{-i\beta}_{0} dt^{'} \mbox{         }
\int_{0}^{-i\beta} dt_{1}
\mbox{      }\theta(t^{'}-t_{1})
\left< T  \mbox{       }{\hat{ {\bf{P}} }}(t_{1}) \cdot 
 {\hat{ {\bf{P}} }}(t) \right>
\label{SIGACZ}
\end{equation}
 Here $ {\bf{ {\hat{P}} }} = \sum_{ {\bf{k}} }{\bf{k}}
 \mbox{      } c^{\dagger}_{ {\bf{k}} }c_{ {\bf{k}} } $.
  At first sight it appears that 
  $ \sigma_{d.c.} $ of Eq.(~\ref{SIGDC}) is not equal to
  $ \sigma_{a.c.}(0) $ of Eq.(~\ref{SIGACZ}). 
 For the two expressions to be the same we must have,
\begin{equation}
{\bf{ {\hat{X}} }}(t^{'})
 = \frac{1}{m} 
\int^{ -i \beta }_{0} dt_{1} \theta(t^{'}-t_{1}) {\bf{ {\hat{P}} }}(t_{1})
\end{equation}
If we take the derivative with respect to $ t^{'} $ we find,
\begin{equation}
\frac{ \partial }{ \partial t }{\bf{ {\hat{X}} }}(t)
 = \frac{1}{m} 
 {\bf{ {\hat{P}} }}(t)
\end{equation}
 This is nothing but the definition of the momentum operator.
 It is a trivial kinematical result. Hence we may conclude that the
 zero frequency limit of the a.c. conductivity is in fact the d.c.
 conductivity.

\end{document}